\documentclass[twocolumn,showpacs,preprintnumbers,showkeys]{revtex4}
\usepackage{graphicx}
\usepackage{dcolumn}
\usepackage{bm}

\begin{document}

\preprint{APS/123-QED}
\title{Hybrid Burnett Equations. \\
A New Method of Stabilizing}
\author{Lars H. S\"{o}derholm}
\affiliation{Mekanik, KTH, SE-100 44 Stockholm, Sweden\\
lars.soderholm@mech.kth.se }
\date{\today }

\begin{abstract}
In the original work by Burnett the pressure tensor and the heat current
contain two time derivates. Those are commonly replaced by spatial
derivatives using the equations to zero order in the Knudsen number. The
resulting conventional Burnett equations were shown by Bobylev to be
linearly unstable. In this paper it is shown that the original equations of
Burnett have a singularity. A hybrid of the original and conventional
equations is constructed which is shown to be linearly stable. It contains
two parameters. For the simplest choice of parameters the hybrid equations
have no third derivative of the temperature but the inertia term contains
second spatial derivatives. For stationary flow, when terms $K\!n^2M\!a^2$
can be neglected, the only difference from the conventional Burnett
equations is the change of coefficients $\varpi _{2}\rightarrow \varpi
_{3},\varpi _{3}\rightarrow \varpi _{3}.$
\end{abstract}

\maketitle

\bigskip

\bigskip

\bigskip

\bigskip

\bigskip


\section{\label{Introduction}Introduction}

The practical interest of extending the Navier-Stokes equations to smaller
length scales comes from the need to model re-entrance of space vehicles and
in later times from nano scale technology. In this parameter region there is
in general the Boltzmann equation. As the collision integral is very heavy
to calculate numerically, the Boltzmann equation is usually applied for
fairly simple or idealized problems, see Cercignani \cite{Cercignanis lilla
bok}. An other method is that of DSMC due to Bird \cite{Bird}. But in the
collision dominated region also DSMC is very demanding computationally.
Hence special methods are valuable when the Knudsen number is too large for
the Navier-Stokes equations to apply but still small enough to allow for an
expansion. A fruitful technique in this region is the asymptotic method,
originating with Hilbert and Grad, see Sone \cite{Sone}.

The Navier-Stokes equations were derived from the Boltzmann equation by the
Chapman-Enskog method, see Chapman \& Cowling \cite{CC}. They are to first
order in the mean free path. The corresponding equations to second order in
the mean free path were derived by Burnett \cite{Burnett}, see also \cite{CC}%
. However, the Burnett equations were proven by Bobylev \cite{Bobylev} to
have a nonphysical instability. See also the review by Agarwal et al \cite%
{Agarwal} and the paper by Uribe et al. \cite{Uribe}.

For the Burnett equations the trivial state of rest is thus unstable for
perturbations of a wavelength of the order of the mean free path and
shorter. The Chapman-Enskog method is an expansion in the Knudsen number $%
K\!n=l/L$, where $l$ is the mean free path and $L$ a characteristic length.
Wavelengths of the order of the mean free path and larger correspond to an
effective Knudsen number of the order $1$ and larger. So there is no
contradiction in the fact that the Burnett equations are unstable for short
wavelengths. The physical content of the Burnett equations is for solutions
with a characteristic length which is large enough. But nevertheless the
equations should be wellbehaved for all length scales. This is necessary
mathematically as well as numerically.

As pointed out by Agarwal et al \cite{Agarwal} Burnett's original expression
for the viscous pressure tensor contains the time derivative of the
traceless rate of deformation tensor. Chapman \&\ Cowling replace the time
derivative by spatial derivatives using the equations at the Euler level.
Agarwal et al give the name conventional Burnett equations to the resulting
equations. There is a similar replacement for the time derivative of the
temperature gradient. This method of replacement using the zero order
equations is part of the procedure in the Chapman-Enskog method to obtain
the Navier-Stokes equations from the Boltzmann equation to first order of
the Knudsen number.

Agarwal et al also raise the question if the instability of the conventional
Burnett equations is caused by this replacement, so that possibly the
original Burnett equations are linearly stable. In the present paper we show
that the original Burnett equations have an unphysical singularity. In \cite%
{Jin} Jin and Slemrod introduce $\mathbf{P,q}$ as new independent fields and
propose 13 equations, first order in time and second order in space. See
also the two master theses by Sv\"{a}rd \cite{Svard} and by Str\"{o}mgren 
\cite{Stromgren} and the paper by S\"{o}derholm \cite{Soderholm} based on
essentially the same idea but with equations which are first order in time
as well as space. In the paper \cite{Soderholm} these equations are applied
numerically for nonlinear sound waves.

In the present paper we introduce a two parameter partial replacement of the
mentioned time derivatives and show that the parameters can be chosen to
give linear stability. We then choose the parameters such that the resulting
equations are as close to the original and conventional equations as
possible. The resulting equations are more similar to the already studied
Burnett equations than those in \cite{Jin} and \cite{Soderholm} and it seems
that it should be simpler to reduce a numerical scheme for the equations to
one for the Navier-Stokes equations.

\section{\label{Burnett}The Burnett Equations}

The general equations of balance are 
\begin{eqnarray}
\frac{D\rho }{Dt}+\rho \mathbf{\nabla \cdot v} &\mathbf{=}&0,
\label{eqn mass} \\
\rho \frac{D\mathbf{v}}{Dt} &=&-\mathbf{\nabla \cdot P,}
\label{eqn momentum} \\
\rho \frac{3k_{B}}{2m}\frac{DT}{Dt} &=&-\mathbf{P}:\mathbf{\nabla v-\nabla
\cdot q.}  \label{egn energy}
\end{eqnarray}%
$\mathbf{P}$ is the pressure tensor and $\mathbf{q}$ the heat current. We
are using dyadic notation.

Let us first write down the original Burnett expression for the pressure
tensor, see \cite{CC} ($p=k_{B}\rho T/m$)

\begin{eqnarray}
\mathbf{P}_o &=&p\mathbf{1}-2\mu \mathbf{S}+\varpi _{1}\frac{\mu ^{2}}{p}(%
\mathbf{\nabla \cdot v})\,\mathbf{S}  \nonumber \\
&&+\varpi _{2}\frac{\mu ^{2}}{p}[\frac{D\mathbf{S}}{Dt}-2\langle \mathbf{S}%
\cdot (\mathbf{\nabla v})\rangle ]  \label{Burnett original P} \\
&&+{\varpi }_{3}\frac{\mu ^{2}}{\rho T}\langle \mathbf{\nabla \nabla }%
T\rangle +{\varpi }_{4}\frac{\mu ^{2}}{\rho pT}\langle \mathbf{\nabla }\,p%
\mathbf{\nabla }\,T\rangle  \nonumber \\
&&+\varpi_{5}\frac{\mu ^{2}}{\rho T^{2}}\langle \mathbf{\nabla }\,T\mathbf{%
\nabla }\,T\rangle +\varpi_{6}\frac{\mu ^{2}}{p}\langle \mathbf{S}\cdot 
\mathbf{S}\rangle .  \nonumber
\end{eqnarray}%
Here,

\begin{eqnarray*}
(\mathbf{\nabla v)}_{ij} &=&\frac{\partial }{\partial x_{i}}v_{j}=v_{j,i},\;(%
\mathbf{\nabla \nabla }T\!)_{ij}=T_{,ij} \\
\mathbf{S} &\mathbf{=}&\langle \mathbf{\nabla v}\rangle =\frac{1}{2}[\mathbf{%
\nabla v+(\nabla v)}^{T}]-\frac{1}{3}(\mathbf{\nabla \cdot v)1}.
\end{eqnarray*}%
$\mathbf{1}$ is the unit tensor, $\langle ...\rangle $ means the symmetric
traceless part. All other quantities have their usual meaning. The original
Burnett expression for the heat current is%
\begin{eqnarray}
\mathbf{q}_o &=&-\kappa \,\mathbf{\nabla }\,\,T+\theta _{1}\frac{\mu ^{2}}{%
\rho T}(\mathbf{\nabla \cdot v)\,\nabla }\,T  \nonumber \\
&&+\theta _{2}\frac{\mu ^{2}}{\rho T}[\frac{D(\mathbf{\nabla }\,T)}{Dt}-%
\mathbf{(\nabla v)\cdot \nabla }\,T]  \label{Burnetts original q} \\
&&+\theta _{3}\frac{\mu ^{2}}{\rho p}\mathbf{S}\cdot \mathbf{\nabla }%
\,\,p+\theta _{4}\frac{\mu ^{2}}{\rho }\mathbf{\nabla }\,\cdot \mathbf{S+}%
\theta _{5}\frac{3\mu ^{2}}{\rho T}\mathbf{S\cdot \nabla }\,T.  \nonumber
\end{eqnarray}

In Chapman \&\ Cowling \cite{CC} the time derivatives $D/Dt$ are replaced by
spatial derivatives using the zero order equations. The resultting
expressions are denoted $D_{0}/Dt.$

To start with we have 
\[
\frac{D}{Dt}(\mathbf{\nabla }\,T)=\mathbf{\nabla }\,\frac{DT}{Dt}-\mathbf{%
(\nabla v)\cdot \nabla }\,T. 
\]%
The energy equation to zero order is 
\[
\frac{3}{2}\frac{D_{0}T}{Dt} =-T(\mathbf{\nabla \cdot v}). 
\]
Hence, 
\begin{eqnarray}
\frac{D_{0}}{Dt}(\mathbf{\nabla }\,T) &&
\label{Burnetts tidsderivata nabla T} \\
=-\frac{2}{3}T\mathbf{\nabla (\nabla \cdot v)-}\frac{2}{3}(\mathbf{\nabla
\cdot v})\mathbf{\nabla }\,T &-&(\mathbf{\nabla v})\cdot \mathbf{\nabla }\,T.
\nonumber
\end{eqnarray}%
Similarly we have 
\begin{eqnarray}
\mathbf{\nabla }\frac{D\mathbf{v}}{Dt} &=&\frac{D}{Dt}\mathbf{\nabla
v+(\nabla v)}^{2},  \nonumber \\
\langle \mathbf{\nabla }\frac{D\mathbf{v}}{Dt}\rangle &=&\frac{D\mathbf{S}}{%
Dt}+\langle \mathbf{(\nabla v)}^{2}\rangle .  \label{DS/Dt och nabla Dv/Dt}
\end{eqnarray}%
The zero order momentum equation is 
\[
\frac{D_{0}\mathbf{v}}{Dt}=-\frac{1}{\rho }\mathbf{\nabla }p. 
\]%
Hence, 
\begin{equation}
\frac{D_{0}\mathbf{S}}{Dt}\mathbf{=}-\langle \mathbf{\nabla (}\frac{1}{\rho }%
\mathbf{\nabla }p)\rangle -\mathbf{\langle \nabla v)}^{2}\rangle .
\label{Burnetts tolkning av tidsderivatan av sparlosa deformationshastighen}
\end{equation}%
Explicitly,%
\begin{eqnarray}
\frac{D_{0}\mathbf{S}}{Dt}+\langle \mathbf{(\nabla v)}^{2}\rangle &&
\label{explicitly} \\
=-\frac{p}{\rho T}\langle \frac{1}{\rho }\mathbf{\nabla }T\mathbf{\nabla }%
\rho -\frac{T}{\rho ^{2}}\mathbf{\nabla }\rho \mathbf{\nabla }\rho &+&\frac{T%
}{\rho }\mathbf{\nabla \nabla }\rho +\mathbf{\nabla \nabla }T\rangle. 
\nonumber
\end{eqnarray}

Interpreting the time derivatives in (\ref{Burnett original P}) and (\ref%
{Burnetts original q} ) according to (\ref{Burnetts tolkning av
tidsderivatan av sparlosa deformationshastighen}) and (\ref{explicitly}) we
find the conventional expressions for the pressure tensor and heat current.
Let us denote them $\mathbf{P_c,q_c.}$ The equations of balance (\ref{eqn
mass}) - (\ref{egn energy}) then give the conventional Burnett equations.

\section{\label{Stability}Stability and Replacing Time derivatives by space
derivatives}

\bigskip We now make the replacements 
\begin{eqnarray}
\frac{D\mathbf{S}}{Dt} &\rightarrow &(1-\alpha )\frac{D\mathbf{S}}{Dt}%
+\alpha \frac{D_{0}\mathbf{S}}{Dt},  \label{alpha replacement} \\
\frac{D}{Dt}(\mathbf{\nabla }\,T) &\rightarrow &(1-\beta )\frac{D}{Dt}(%
\mathbf{\nabla }\,T)+\beta \frac{D_{0}}{Dt}(\mathbf{\nabla }\,T),
\label{beta replacement}
\end{eqnarray}%
where $\alpha ,\beta $ are coefficients for which we later shall obtain bounds. The choice $%
\alpha =\beta =0$ gives the original Burnett equations and $\alpha =\beta =1$
the conventional Burnett equations. Let us now for simplicity denote
nonlinear Burnett terms by dots.%
\[
\mathbf{P}=p\mathbf{1}-2\mu \mathbf{S}+\varpi _{2}\frac{\mu ^{2}}{p}\langle 
\mathbf{\nabla }\frac{D\mathbf{v}}{Dt}\rangle +\varpi_{3}\frac{\mu ^{2}}{%
\rho T}\langle \mathbf{\nabla \nabla }T\rangle +... \nonumber
\]%
This gives%
\begin{eqnarray}
\mathbf{P} &=&p\mathbf{1}-2\mu \mathbf{S}+\varpi _{2}\frac{\mu ^{2}}{p}%
(1-\alpha )\langle \mathbf{\nabla }\frac{D\mathbf{v}}{Dt}\rangle \\
&+&\frac{\mu ^{2}}{\rho T}(\varpi_{3}-\alpha \varpi _{2})\langle \mathbf{%
\nabla \nabla }T\rangle -\alpha \varpi _{2}\frac{\mu ^{2}}{\rho ^{2}}\langle%
\mathbf{\nabla \nabla }\rho \rangle +...  \nonumber
\end{eqnarray}

In order to study the linear stability of the resulting equations we now
linearize the equations around a state at rest with constant temperature and
density. 
\begin{eqnarray*}
&&\rho \lbrack 1+\varpi _{2}\frac{\mu ^{2}}{p\rho }(1-\alpha )(\frac{1}{2}%
\triangle +\frac{1}{6}\mathbf{\nabla \nabla )}]\frac{\partial \mathbf{v}}{%
\partial t} \\
&=&-\frac{p}{\rho }(1-\alpha \varpi _{2}\frac{\mu ^{2}}{\rho p}\frac{2}{3}%
\triangle )\mathbf{\nabla }\rho \\
&-&\frac{p}{T}\lbrack1+\frac{\mu ^{2}}{\rho p}(\varpi_{3}-\alpha \varpi _{2})%
\frac{2}{3}\triangle \rbrack\mathbf{\nabla }T+\mu \lbrack \triangle \mathbf{%
v+}\frac{1}{3}\mathbf{\nabla (\nabla \cdot v)}].
\end{eqnarray*}%
All the undifferentiated quantities are here taken at the background state.
For a plane wave with wave number $k$ the longitudinal component of the
momentum equation gives 
\begin{eqnarray*}
&&\frac{\partial v_{\Vert }}{\partial t} \\
&=&-\frac{1+\alpha \varpi _{2}\frac{\mu ^{2}}{\rho p}\frac{2}{3}k^{2}}{%
1+\varpi _{2}\frac{\mu ^{2}}{p\rho }(\alpha -1)\frac{2}{3}k^{2}}\frac{k_{B}T%
}{m\rho }\mathbf{\nabla }_{\Vert }\rho \\
&&-\frac{1+\frac{\mu ^{2}}{\rho p}(\alpha \varpi _{2}-\varpi_{3}) \frac{2}{3}%
k^{2}}{1+\varpi _{2}\frac{\mu ^{2}}{p\rho }(\alpha -1)\frac{2}{3}k^{2}}\frac{%
k_{B}}{m}\mathbf{\nabla }_{\Vert }T \\
&&+\frac{1}{1+\varpi _{2}\frac{\mu ^{2}}{p\rho }(\alpha -1)\frac{2}{3}k^{2}}%
\frac{4\mu }{3\rho }\triangle v_{\Vert }
\end{eqnarray*}

Putting $k=0$ here we have the linearized Navier-Stokes equations. To avoid
a singularity in the coefficients for a general $k$ we see that it is
necessary that $\alpha \geq 1$. As $\alpha=0$ corresponds to the original
Burnett equations, we see that they have an unphysical singularity. - For a
given $k$ we can interpret the equations as the linearized Navier-Stokes
equations for an ideal gas but with different properties of the gas and the
background state 
\begin{eqnarray*}
\frac{\hat{T}}{\hat{m}} & = &\frac{T}{m}\frac{1+\alpha \varpi _{2}\frac{\mu
^{2}}{\rho p}\frac{2}{3}k^{2}}{1+\varpi _{2}\frac{\mu ^{2}}{p\rho }(\alpha
-1)\frac{2}{3}k^{2}}, \\
\frac{1}{\hat{m}} & = &\frac{1}{m}\frac{1+\frac{\mu ^{2}}{\rho p}(\alpha
\varpi _{2}-\varpi_{3})\frac{2}{3}k^{2}}{1+\varpi _{2}\frac{\mu ^{2}}{p\rho }%
(\alpha -1)\frac{2}{3}k^{2}}, \\
\hat{\mu} & = &\mu \frac{1}{[1+\varpi _{2}\frac{\mu ^{2}}{p\rho }(\alpha -1)%
\frac{2}{3}k^{2}]}.
\end{eqnarray*}
This interpretation is possible if the coefficients are positive. We see
that this requires 
\[
\alpha \varpi _{2}-\varpi_{3}\geq 0. 
\]%
As $0<\varpi _{2}<\varpi_{3}$ for Maxwell molecules as well as hard spheres, 
$\alpha $ then has to be larger than $1$, which excludes the original
Burnett equations as well as the conventional ones.

Using 
\[
\mathbf{\nabla }\,\cdot \mathbf{S=}\frac{1}{2}\triangle \mathbf{v+}\frac{1}{6%
}\mathbf{\nabla (\nabla \cdot v)} 
\]%
we obtain for the heat current 
\begin{eqnarray*}
\mathbf{q} &=&-\kappa \,\mathbf{\nabla }\,\,T+\theta _{2}\frac{\mu ^{2}}{%
\rho T}(1-\beta )\frac{D(\mathbf{\nabla }\,T)}{Dt} \\
&&+\frac{\mu ^{2}}{\rho }[(\frac{\theta _{4}}{2})\triangle \mathbf{v}+(\frac{%
\theta _{4}}{6}-\beta\frac{2\theta _{2}}{3})\mathbf{\nabla (\nabla
\cdot v)]+...}
\end{eqnarray*}%
The linearized energy equation is then%
\begin{eqnarray}
&&\frac{3k_B\rho}{2m}(1+\theta _{2}\frac{2}{3}\frac{\mu ^{2}}{\rho p}%
(1-\beta )\triangle )\frac{\partial T}{\partial t}  \nonumber \\
&=&-p(1+\frac{\mu ^{2}}{\rho p}\frac{2}{3}(\theta _{4}-\beta\theta
_{2})\triangle )(\mathbf{\nabla \cdot v)+}\kappa \,\triangle \,\,T.
\end{eqnarray}

For a plane wave we obtain 
\begin{eqnarray*}
&&\frac{\partial T}{\partial t} =-\frac{T}{mc_{v}}\frac{1+\frac{\mu ^{2}}{%
\rho p}\frac{2}{3}(\beta\theta _{2}-\theta _{4})k^{2}}{1+\theta
_{2}\frac{2}{3}\frac{\mu ^{2}}{\rho p}(\beta -1)k^{2}}(\mathbf{\nabla \cdot
v)} \\
&&\mathbf{+}\frac{\kappa }{\rho c_{v}}\frac{1}{1+\theta _{2}\frac{2}{3}\frac{%
\mu ^{2}}{\rho p}(\beta -1)k^{2}}\,\triangle \,\,T.
\end{eqnarray*}%
Here we have introduced the specific heat%
\[
c_{v}=\frac{3k_{B}}{2m}. 
\]%
To avoid singularities it is necessary that $\beta \geq 1$. This means that
the conventional Burnett equations have no singularity in the energy
equation, but the original Burnett equations do. We can interpret the energy
equation as the energy equation for the Navier-Stokes equations of an ideal
gas with Fourier's expression for the heat current. 
\begin{eqnarray*}
\frac{\hat{T}}{\hat{m} \hat{c_{v}}} & = &\frac{T}{m c_{v}}\frac{1+\frac{\mu
^{2}}{\rho p}\frac{2}{3}(\beta\theta _{2}-\theta _{4})k^{2}}{%
1+\theta _{2}\frac{2}{3}\frac{\mu ^{2}}{\rho p}(\beta -1)k^{2}}, \\
\frac{\hat{\kappa}}{\hat{c_v}} &= &\frac{\kappa}{c_v} \frac{1}{1+\theta
_{2}\frac{2}{3}\frac{\mu ^{2}}{\rho p}(\beta -1)k^{2}}.
\end{eqnarray*}

In all we have then given new formal values to the five properties of the
gas and its background state $m, c_v, \mu, \kappa, T$. $\rho$ and $\mathbf{v}
$ are unchanged. This interpretation is possible as long as the coefficients
are positive. As $0<\theta _{4}<\theta _{2}$ for Maxwell molecules as well
as hard spheres this follows from $\beta \geq 1.$ As the linearized equation
of continuity is the usual one we conclude that for each value $k$ there is
an ideal gas such that its linearized Navier-Stokes equations coincide with
the full set of linearized hybrid Burnett equations for the same value of $k$%
. This means that the equations are linearly stable as long as 
\begin{equation}
\alpha \geq \frac{\varpi_{3}}{\varpi_{2}},\beta \geq 1.
\end{equation}
So we have found a two-parameter family of equations with linear stability.
In this way of thinking, it is easy to understand that the conventional
Burnett equations are linearly unstable. For large enough $k$ a temperature
gradient gives rise to a pressure gradient in the opposite direction. So if
a local region of the order of the mean free path or smaller is hotter than
its surroundings, the pressure in this area will be lower and the gas will
flow into it, increasing its temperature. But let us stress that this is for
length scales of the order of the mean free path and smaller, so that the
equations are not physically valid anyhow. As mentioned in the introduction,
the equations nevertheless need to be well-behaved for such perturbations.

For the transverse components we find

\begin{eqnarray*}
&&\rho \lbrack 1+\varpi _{2}\frac{\mu ^{2}}{p\rho }(\alpha -1)\frac{1}{2}%
k^{2}]\frac{\partial \mathbf{v}_{\mathbf{\bot }}}{\partial t} \\
&=&\mu \triangle \mathbf{v}_{\mathbf{\bot }}
\end{eqnarray*}%
As long as $\alpha \geq 1$ this is simply diffusion, with a diffusivity
depending on $k$.

\section{\label{Hybrid equations}Hybrid Burnett Equations}

Let us now make the following choice of the parameters

\begin{equation}
\alpha =\frac{\varpi _{3}}{\varpi _{2}},\beta =1.
\end{equation}%
The heat current then is the conventional one, $\mathbf{q}_{c}$. The
resulting expression for the viscous pressure tensor is now denoted $\mathbf{%
P}_{h}$. 
\begin{eqnarray}
&&\mathbf{P}_{h}=p\mathbf{1}-2\mu \mathbf{S}+\varpi _{1}\frac{\mu ^{2}}{p}(%
\mathbf{\nabla }\cdot \mathbf{v})\,\mathbf{S}  \nonumber \\
&&+(\varpi _{2}-\varpi _{3})\frac{\mu ^{2}}{p}\frac{D\mathbf{S}}{Dt} 
\nonumber \\
&&-2\varpi _{2}\frac{\mu ^{2}}{p}\langle \mathbf{S}\cdot (\mathbf{\nabla v}%
)\rangle -\varpi _{3}\frac{\mu ^{2}}{\rho ^{2}}\langle \mathbf{\nabla \nabla 
}\rho \rangle   \nonumber \\
&&+\varpi _{3}\frac{\mu ^{2}}{\rho T}\langle -\frac{1}{\rho }\mathbf{\nabla }%
T\mathbf{\nabla }\rho +\frac{T}{\rho ^{2}}\mathbf{\nabla }\rho \mathbf{%
\nabla }\rho -\frac{p}{\rho T}(\mathbf{\nabla v)}^{2}\rangle   \nonumber \\
&&+\varpi _{4}\frac{\mu ^{2}}{\rho pT}\langle \mathbf{\nabla }\,p\mathbf{%
\nabla }\,T\rangle +\varpi _{5}\frac{\mu ^{2}}{\rho T^{2}}\langle \mathbf{%
\nabla }\,T\mathbf{\nabla }\,T\rangle   \nonumber \\
&&+\varpi _{6}\frac{\mu ^{2}}{p}\langle \mathbf{S}\cdot \mathbf{S}\rangle . 
\nonumber
\end{eqnarray}%
Note that the troublesome third derivative of $T$ is absent. Further there
is a change of sign in front of $D\mathbf{S/}Dt$ as compared to the original
expression (\ref{Burnett original P}). 

Now we have our hybrid Burnett equations. 
\begin{eqnarray}
\frac{D\rho }{Dt}+\rho \mathbf{\nabla \cdot v} &\mathbf{=}&0, \\
\rho \frac{D\mathbf{v}}{Dt} &=&-\mathbf{\nabla \cdot P}_{h}, 
\label{hybrid momentum} \\
\rho \frac{3k_{B}}{2m}\frac{DT}{Dt} &=&-\mathbf{P}_{h}:\mathbf{\nabla
v-\nabla \cdot q}_{c}\mathbf{.}
\end{eqnarray}

\section{\label{Linear stability}Linear stability analysis}

Let us linearize around a uniform state at rest denoting now its
temperature $T_0$ and density $\rho_0$. We write 
\[
T=T_{0}(1+\tilde{T}),\;\rho =\rho _{0}(1+\tilde{\rho}),\;v=\sqrt{\frac{%
k_{B}T_{0}}{m}}\tilde{v}. 
\]%
We introduce the dimensionless variables, where the unit of length is of the
order of the mean free path

\[
x=x^{\ast }\frac{\mu _{0}}{\rho _{0}}\sqrt{\frac{m}{k_{B}T_{0}}},\;t=t^{\ast
}\frac{\mu _{0}}{\rho _{0}}\frac{m}{k_{B}T_{0}}. 
\]%
In the rest of this section stars and tildes are omitted and subscripts
denote partial derivatives. We obtain in the longitudinal case 
\begin{eqnarray*}
\rho _{t} +v_{x} =0 \\
\rho _{x}-\frac{2}{3}\varpi _{3}\rho _{xxx} \\
+v_{t}-\frac{4}{3}v_{xx}-\frac{2}{3}(\varpi _{3}-\varpi _{2})v_{xxt} +T_{x}
=0 \\
\frac{2}{3} v_{x}+\frac{4}{9}(\theta _{4}-\theta _{2})v_{xxx}+T_t -fT_{xx} =0
\end{eqnarray*}%
$f=(2m\kappa) /(3k_{B}\mu) =5/(3 P\!r )$, where $P\!r $ is the Prandtl
number. Looking for solutions $\exp (ikx+\Lambda t)$, we find the determinant

\begin{eqnarray}
&&\Lambda ^{3}(1+\frac{2\varpi _{3}}{3}k^{2})+\Lambda ^{2}[\frac{2}{3}%
(\varpi _{3}-\varpi _{2})fk^{4}+(\frac{4}{3}+f)k^{2}]  \nonumber \\
&&+\Lambda \lbrack (\frac{2\varpi _{3}}{3}-\frac{4}{9}(\theta _{4}-\theta
_{2}))k^{4}+\frac{5}{3}k^{2}]+(\varpi _{3}fk^{6}+fk^{4}).  \nonumber
\end{eqnarray}

Let us in particular consider the asymptotic case $k\rightarrow \infty $. We
find one mode with 
\[
\Lambda _{0}\approx -\frac{(\varpi _{3}-\varpi _{2})f}{\varpi _{3}}k^{2}. 
\]%
There are also two complex conjugate roots%
\begin{eqnarray*}
\Lambda _{\pm }\approx \pm i\sqrt{\frac{3\varpi _{3}}{2(\varpi _{3}-\varpi
_{2})}}k \\
-\frac{9\varpi _{3}^{2}+2(\varpi _{3}-\varpi _{2})[3\varpi _{3}+2(\theta
_{2}-\theta _{4})]}{6(\varpi _{3}-\varpi _{2})^{2}f}.
\end{eqnarray*}

For Maxwell molecules, see \cite{CC} 
\[
\theta _{2}=45/8,\theta _{4}=3,\varpi _{2}=2,\varpi _{3}=3. 
\]%
For hard spheres, see \cite{Reinecke & Kremer}

\[
\theta _{2}=5.826,\theta _{4}=2.416,\varpi _{2}=2.029,\varpi _{3}=2.415. 
\]%
In both cases $\varpi _{3},\varpi _{3}-\varpi _{2},\theta _{2}-\theta _{4}$
are positive. This means that the mode corresponding to $\Lambda _{0}$ is
damped and nonpropagating. It is the entropy mode. The two modes $\Lambda
_{\pm }$ are damped. They are propagating sound waves.

We plot the three roots $\Lambda $\ in the complex plane in Fig. 1. Here we
use the value the Prandtl number $2/3$. This is the lowest approximation in
terms of Sonine polynomial expansion for any interatomic potential and is
experimentally found to be a good approximation, see \cite{CC}. Prandtl
numbers in the range between $1/2$ to $1$ give qualitatively the same plot
as Fig. 1. We see in Fig. 1 that the real part of $\Lambda $ is negativ.

\begin{figure}
\includegraphics[width=8cm]{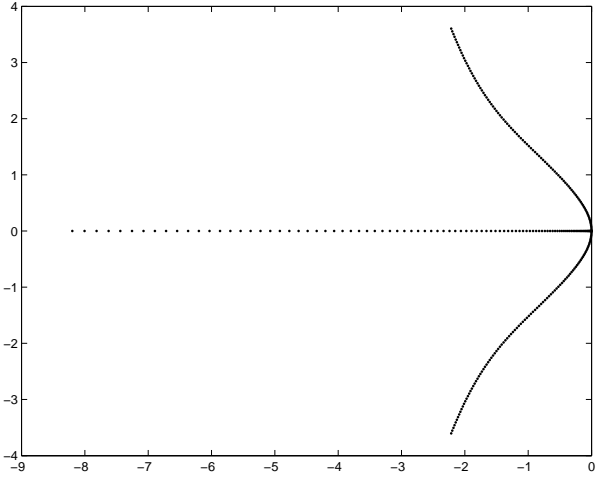}
\caption{Complex growth factor $\Lambda $ for $0\leq
k\leq 2$. Hard spheres.}
\end{figure}

We have already in the preceding section found that the transverse mode is
damped and nonpropagating.

\section{\label{Discussion}Discussion}

Let us pull out the acceleration from the viscous pressure tensor $\mathbf{P}_{h}$ using 
(\ref{DS/Dt och nabla Dv/Dt}
)%
\[
\mathbf{P}_{h}=(\varpi _{2}-\varpi _{3})\frac{\mu ^{2}}{p}\langle (\mathbf{%
\nabla }\frac{D\mathbf{v}}{Dt})\rangle +\mathbf{\tilde{P}}_{h}.
\]%
Here
\begin{eqnarray}
&&\mathbf{\tilde{P}}_{h}=p\mathbf{1}-2\mu \mathbf{S}+\varpi _{1}\frac{\mu
^{2}}{p}(\mathbf{\nabla }\cdot \mathbf{v})\,\mathbf{S}  \nonumber \\
&&-(\varpi _{2}-\varpi _{3})\frac{\mu ^{2}}{p}\langle (\mathbf{\nabla v}%
)^{2}\rangle   \nonumber \\
&&-2\varpi _{2}\frac{\mu ^{2}}{p}\langle \mathbf{S}\cdot (\mathbf{\nabla v}%
)\rangle -\varpi _{3}\frac{\mu ^{2}}{\rho ^{2}}\langle \mathbf{\nabla \nabla 
}\rho \rangle   \nonumber \\
&&+\varpi _{3}\frac{\mu ^{2}}{\rho T}\langle -\frac{1}{\rho }\mathbf{\nabla }%
T\mathbf{\nabla }\rho +\frac{T}{\rho ^{2}}\mathbf{\nabla }\rho \mathbf{%
\nabla }\rho -\frac{p}{\rho T}(\mathbf{\nabla v)}^{2}\rangle   \nonumber \\
&&+\varpi _{4}\frac{\mu ^{2}}{\rho pT}\langle \mathbf{\nabla }\,p\mathbf{%
\nabla }\,T\rangle +\varpi _{5}\frac{\mu ^{2}}{\rho T^{2}}\langle \mathbf{%
\nabla }\,T\mathbf{\nabla }\,T\rangle   \nonumber \\
&&+\varpi _{6}\frac{\mu ^{2}}{p}\langle \mathbf{S}\cdot \mathbf{S}\rangle . 
\nonumber
\end{eqnarray}%
The momentum equation can now be written
\[
\rho \frac{D\mathbf{v}}{Dt}-\mathbf{\nabla }\cdot \lbrack (\varpi
_{3}-\varpi _{2})\frac{\mu ^{2}}{p}\langle (\mathbf{\nabla }\frac{D\mathbf{v}%
}{Dt})\rangle ]=-\mathbf{\nabla \cdot \tilde{P}}_{h}.
\]%
Let us write $\mathbf{a}$ for the acceleration and study the terms
containing the acceleration 
\[
\rho \mathbf{a}-\mathbf{\nabla }\cdot \lbrack (\varpi _{3}-\varpi _{2})\frac{%
\mu ^{2}}{p}\langle \mathbf{\nabla a}\rangle ]
\]%
We multiply by $\mathbf{a}$ and integrate over the region of flow%
\begin{eqnarray*}
&&\int \{\mathbf{a}\cdot [\rho \mathbf{a}-\mathbf{\nabla }\cdot ((\varpi
_{3}-\varpi _{2})\frac{\mu ^{2}}{p}\langle \mathbf{\nabla a}\rangle )]\}dV \\
&=&\int [\rho a^{2}+(\varpi _{3}-\varpi _{2})\frac{\mu ^{2}}{p}\langle 
\mathbf{\nabla a}\rangle :\langle \mathbf{\nabla a}\rangle ]dV\\&-&\int (\varpi
_{3}-\varpi _{2})\frac{\mu ^{2}}{p}\mathbf{a}\cdot \langle \mathbf{\nabla a}%
\rangle \cdot d\mathbf{S}
\end{eqnarray*}%
If there are no boundaries and the flow vanishes at infinity, the surface
integral vanishes. We conclude that the operator acting on $D\mathbf{v}/Dt$
is then positive and the momentum equation can be solved for the acceleration.
The situation when there are boundaries requires a closer examination.  

In order to see more clearly the structure of the hybrid Burnett equations,
we replace the nonlinear terms in $K\!n^{2}$ by dots. But first we split up
in longitudinal and transverse parts according to%
\[
\frac{1}{2}\triangle +\frac{1}{6}\mathbf{\nabla \nabla }=\frac{2}{3}\mathbf{%
\nabla \nabla }+\frac{1}{2}(\triangle -\mathbf{\nabla \nabla }).\label{long
and transverse second derivative}
\]
The hybrid Burnett equations can then be written
\begin{eqnarray}
\frac{D\rho }{Dt}+\rho (\mathbf{\nabla \cdot v}) =0,&&
\label{mass, Burnett linear} \\
\rho \{\mathbf{1}-(\varpi _{3}-\varpi _{2})\frac{\mu ^{2}}{p\rho }[\frac{1}{2%
}(\triangle \mathbf{1-\nabla \nabla }) +\frac{2}{3}\mathbf{\nabla \nabla }%
\}\cdot\frac{D\mathbf{v}}{Dt}&&  \label{momentum Burnett linear} \\
=-\mathbf{\nabla }p+2\mathbf{\nabla} \cdot (\mu \mathbf{S})+\varpi _{3}\frac{%
\mu ^{2}}{\rho ^{2}}\frac{2}{3}\triangle\mathbf{\nabla } \rho +...,&& 
\nonumber \\
\frac{3k_{B}}{2m}\rho \frac{DT}{Dt} =-p(\mathbf{\nabla} \cdot \mathbf{v})
+2\mu \mathbf{S}\cdot\mathbf{S}+\mathbf{\nabla} \cdot (\kappa \,\mathbf{%
\nabla }\,\,T)&& \\
-(\theta _{4}-\theta _{2})\frac{\mu ^{2}}{\rho }\frac{2}{3}\triangle (%
\mathbf{\nabla} \cdot \mathbf{v})+...&&  \nonumber
\end{eqnarray}

Let us also write down the momentum equation of the conventional Burnett
equations There is no need to write down the equations of continuity and
energy as they are exactly the same as for the hybrid equations. 
\begin{eqnarray}
\rho \frac{D\mathbf{v}}{Dt} =-\mathbf{\nabla }p+2\mathbf{\nabla} \cdot (\mu 
\mathbf{S})-(\varpi _{3}-\varpi _{2})\frac{\mu ^{2}}{\rho T}\frac{2}{3}
\triangle\mathbf{\nabla } T&& \\
+\varpi _{2}\frac{\mu ^{2}}{\rho ^{2}}\frac{2}{3}\triangle\mathbf{\nabla }%
\rho +...&&  \nonumber
\end{eqnarray}%
The difference is that the hybrid Burnett equations have a more complicated
inertia term but no term $\triangle \mathbf{\nabla}T$ in the momentum
equation. The coefficient in front of the term $\triangle\mathbf{\nabla }\rho
$ is $\varpi _{3}-\varpi _{2}$ in the conventional Burnett equations but $%
\varpi _{3}$ in the hybrid Burnett equations.

Let us also note that the equations without the $K\!n^{2}$ nonlinear terms
are valid when $K\!n^{2} M\!a^2$ can be neglected and the relative
temperature and density variations are assumed to be of the order of $M\!a$.

Let us take a closer look at the inertia term in the hybrid Burnett
equations, see (\ref{hybrid momentum}). We consider a Fourier component of
the acceleration ${D\mathbf{v}}/{Dt}$ proportional to $\exp (i\mathbf{k}\cdot%
\mathbf{r})$. Then 
\begin{eqnarray*}
&&\rho \{[\mathbf{1}-(\varpi _{3}-\varpi _{2})\frac{\mu ^{2}}{p\rho }[\frac{1%
}{2}(\triangle \mathbf{1-\nabla \nabla })\mathbf{+}\frac{2}{3}\mathbf{\nabla
\nabla }]\}\frac{D\mathbf{v}}{Dt}  \nonumber \\
&=&\rho \{[1+\frac{1}{2}(\varpi _{3}-\varpi _{2})\frac{\mu ^{2}k^{2}}{p\rho }%
](\mathbf{1}-\frac{\mathbf{kk}}{k^{2}}) \\
&&+[1+\frac{2}{3}(\varpi _{3}-\varpi _{2})\frac{\mu ^{2}k^{2}}{p\rho }]\frac{%
\mathbf{kk}}{k^{2}}\}\frac{D\mathbf{v}}{Dt}.
\end{eqnarray*}%
We have effectively a transverse inertia in the first term and a
longitudinal inertia in the second term. As $\varpi _{3}-\varpi _{2}>0$,
both of them are positive. We noted earlier that the sign of the $D\mathbf{v/%
}Dt$ term of the hybrid Burnett expression for $\mathbf{P}$ is the opposite
of that of the original Burnett equations. As a result the original Burnett
equations have a singularity for certain wave numbers where the inertia
vanishes.

Let us also consider the low $M\!a$ stationary case. The hybrid Burnett
equations are then%
\begin{eqnarray*}
\mathbf{\nabla \cdot (}\rho \mathbf{v}) &=&0, \\
\rho (\mathbf{v\cdot \nabla })v =-\mathbf{\nabla }p&+&2\mathbf{\nabla \cdot (%
}\mu \mathbf{S})+\varpi _{3}\frac{\mu ^{2}}{\rho ^{2}}\frac{2}{3} \triangle%
\mathbf{\nabla }\rho \mathbf{,} \\
\frac{3k_{B}}{2m}\rho (\mathbf{v\cdot \nabla }T) &=&-p\mathbf{(\nabla \cdot
v)+}2\mu \mathbf{S}:\mathbf{S} \\
+\mathbf{\nabla} \cdot (\kappa \,\mathbf{\nabla}\,T) &-&(\theta _{4}-\theta
_{2})\frac{\mu ^{2}}{\rho }\frac{2}{3}\triangle (\mathbf{\nabla} \cdot 
\mathbf{v}).  \nonumber
\end{eqnarray*}%
The conventional Burnett momentum equation is%
\begin{eqnarray}
&&\rho (\mathbf{v\cdot \nabla )v} =-\mathbf{\nabla }p+2\mathbf{\nabla \cdot (%
}\mu \mathbf{S}) \\
&&-(\varpi _{3}-\varpi _{2}\mathbf{)}\frac{\mu ^{2}}{\rho T}\frac{2}{3}
\triangle\mathbf{\nabla } T +\varpi _{2}\frac{\mu ^{2}}{\rho ^{2}}\frac{2}{3}
\triangle\mathbf{\nabla }\rho \mathbf{,}  \nonumber
\end{eqnarray}%
The only difference between the hybrid Burnett equations and the
conventional Burnett equations is the change of coefficients 
\begin{equation}
\varpi _{2}\rightarrow \varpi _{3},\varpi _{3}\rightarrow \varpi _{3}.
\label{only difference mine Burnett stationary low velocity}
\end{equation}
This also means that the third derivative of the temperature is absent in
the hybrid Burnett equations.

\begin{acknowledgments}
During the years I have had many stimulating discussions with Prof. Y. Sone
on the relation between the asymptotic method and the Chapman-Enskog
expansion.
\end{acknowledgments}

\end{document}